# Excitation of spin waves by tunnelling electrons in one-dimensional ferro- and anti-ferromagnets

## J.P. Gauyacq[1,2] and N.Lorente[3]

[1]CNRS, Institut des Sciences Moléculaires d'Orsay, ISMO, UMR 8214, Bâtiment 351, 91405 Orsay Cedex, France
[2]Université Paris-Sud, Institut des Sciences Moléculaires d'Orsay, ISMO, UMR 8214, Bâtiment 351, 91405 Orsay Cedex, France
[3] Centre d'Investigació en Nanociéncia i Nanotecnologia (CSIC-ICN), Campus de la UAB, E-08193 Bellaterra, Spain

## Abstract

Excitation of finite chains of magnetic atoms adsorbed on a surface by tunnelling electrons from an STM (Scanning Tunnelling Microscope) tip is studied using a Heisenberg Hamiltonian description of the magnetic couplings along the chain and a strong coupling approach of inelastic tunnelling. The excitation probability of the magnetic levels is very high and the excitation spectra in chains of different lengths are very similar. The excitations in finite chains can be considered as spin waves quantized in the finite object. The energy and momentum spectra of the spin waves excited in the idealized infinite chain by tunnelling electrons are determined from the results on the finite chains. Both ferro- and anti-ferromagnetic couplings are considered, leading to very different results. In particular, in the anti-ferromagnetic case, excitations linked to the entanglement of the chain ground state are evidenced.

# 1. Introduction

The development of low-temperature Inelastic Electron Tunnelling Spectroscopy (IETS) made possible detailed studies of single adsorbates on metal surfaces and in particular studies of their magnetic properties[1,2,3,4,5,6,7,8]. It was shown that single adsorbates on metal surfaces with an ultra-thin insulator coating can carry a local spin which interacts with the adsorbate surroundings. This interaction leads to a magnetic anisotropy of the local spin, i.e. to the existence of several magnetic energy levels of the adsorbate corresponding to the relative orientation of the local spin with respect to the substrate. These experimental studies also revealed that tunnelling electrons were extremely efficient in inducing transitions between these local magnetic states or equivalently that electron transport through these adsorbates could be dominated by inelastic effects. In the case of adsorbed metal-phthalocyanine molecules, experiments reported inelastic currents three times larger than the elastic current[5]. These results are at variance with the case of vibrational excitation of adsorbates by tunnelling electrons, where the inelastic currents are at most in the few per cent range of the elastic current, as observed in experimental vibrational IETS[9,10] and predicted theoretically[11,12]. A series of theoretical studies was devoted to the magnetic excitation problem in the case of single adsorbates, both to the origin of the magnetic anisotropy[13,14] and to the description of inelastic tunnelling using perturbative[15,16,17,18,19] and strong coupling approaches[20,21]. In particular, the strong coupling approach yielded a physical view of the excitation mechanism and a detailed account of the large efficiency of tunnelling electrons in inducing inelastic effects, as well as an account of the finite lifetime of magnetic excitations due to electron-hole pair creation[22]; this also brought forward the link between magnetic excitation and other angular momentum transfer processes in other surface and molecular physics problems[23,24,25,26].

Tunnelling electrons should also be extremely efficient in inducing magnetic excitations in the case of a magnetic atom lattice, i.e. in the case of local spins coupled together via a ferromagnetic or an anti-ferromagnetic interaction. Several recent experimental studies have been devoted to such systems in one dimension systems of finite size[3,5]; they showed that indeed tunnelling electrons were efficiently inducing magnetic transitions in finite size systems, in a way similar to the case of individual adsorbates. The corresponding energy spectrum of the magnetic excitations could be resolved by IETS and a detailed spectroscopic analysis of the magnetic excitation energies showed that a linear chain of Mn magnetic atoms or of CoPc molecules could efficiently be represented by a Heisenberg

Hamiltonian possibly with magnetic anisotropy terms for each atom[3,5]. In addition, it was shown that the excitation probability of these magnetic excitations was very large for an STM tip positioned above one of the atoms. On the theoretical side, a perturbative approach has been applied to the case of small anti-ferromagnetic Mn chains (1-4 atoms)[16]; it confirmed the qualitative difference between the inelastic conductance of chains with even and odd numbers of atoms, as well as the dependence of the excitation on the tip position along the chain. In the case of an infinite system, magnetic excitations would correspond to spin waves propagating along the lattice. Spin waves have been the subject of many studies in particular via their excitation by neutron scattering[27,28,29,30]. In neutron scattering, the excitation concerns the entire chain and the corresponding dynamic spin structure factor has been described in a series of papers (see e.g.[31,32,33,34] and references therein). In contrast to the neutron scattering case, for tunnelling electrons, the primary excitation is local, the electron tunnelling through a given atom in the chain, which thus 'receives' the excitation; however, this atom is magnetically coupled to the rest of the chain so that the local excitation results in magnetic excitations delocalized along the chain. In the case of excitation by tunnelling electrons coming from an STM tip, experimental evidence of spin wave excitation has been brought in Ref.[35,36,37]. In addition, one could also consider the experimental studies mentioned above on finite size 1D-systems[3,5] as studies of spin wave excitations; in that case, one would consider that the spin waves are quantized by the finite size of the system and that the collective magnetic excitations observed in finite size chains of atoms can be seen as quantized spin waves.

In the present work, we extend our strong coupling treatment of the excitation of local spins in single adsorbates by tunnelling electrons (see e.g. in Ref.[20,21]) to the case of a finite size set of magnetic atoms coupled by magnetic exchange interactions. We chose a highly quantal model system, a set of spin ½; it is of model character although it bears some resemblance with the CoPc anti-ferromagnetic system experimentally studied by Chen et al[5]. Other systems studied by neutron scattering[31,34] also correspond to 1D-chains of spin-1/2. Both ferromagnetic and anti-ferromagnetic interactions are considered and are shown to lead to very different excitation properties. We study finite size systems in one dimension, i.e. linear chains or closed rings of atoms. Besides the extension of the 'strong coupling' theoretical approach and the characterization of magnetic excitations in finite ensembles of coupled local spins, the emphasis of the present work is put on how the excitation of a finite size chain of atoms can be described in terms of the spin waves of the idealized infinite 1D-chain. In particular, we determine the energy and momentum spectra of the spin waves excited by tunnelling electrons in the infinite idealized chain.

## 2. Method

### 2.1 Spectroscopy of the studied systems

We study a finite size ensemble of local spins (a chain of N atoms with S= ½) coupled by a Heisenberg Hamiltonian:

$$H = \sum_{i=1}^{N-1} J\, \vec{S}_i \; \vec{S}_{i+1} \qquad (1)$$

In the present model study, we chose a simple system: the exchange coupling, $J$, is constant and only concerns first neighbours. We also did not include an anisotropy term that would couple each local spin to the substrate orientation. These two approximations can be lifted at will and would not lead to any difficulty in the treatment; we do not think they would influence the qualitative conclusions of the present model study. Two different kinds of systems are examined: open linear chains of atoms (with Hamiltonian (1)) and rings of atoms with Hamiltonian (2):

$$H = \sum_{i=1}^{N} J\, \vec{S}_i \cdot \vec{S}_{i+1} \qquad , with\; \vec{S}_{N+1} \equiv \vec{S}_1 \qquad (2)$$

The latter, with the loop condition, is an attempt to describe an infinite periodic chain of atoms[39]. Both the cases of $J < 0$ and $J > 0$ (ferro and anti-ferromagnets) are studied with more emphasis on the more complex system of anti-ferromagnets. Below we mainly report on even numbers of atoms in the chains and rings; in a ferromagnet, this does not play a role whereas in the anti-ferromagnet case, an even number of atoms ensures that the ground state of the system is a zero spin system, better suited for the extension to infinite chains. As seen below, the exchange interaction, $J$, appears as the energy scale of the magnetic structure in the system and below, $J$ will be used as the energy unit when presenting the results of our study. The use of Heisenberg Hamiltonian to describe a magnetic structure can be questioned in particular for itinerant magnetism but should be appropriate for the case of magnetic moments well-localized on atomic sites (see e.g. a recent discussion and spin wave treatment in [38] and references therein)

Hamiltonian (1) and (2) are diagonalized in a basis of products of local spins of the form $\left| M_1, M_2, \ldots, M_{N-1}, M_N \right\rangle$, where $M_i$ is the projection of the spin of the atom at site $i$ on the quantization axis ($M_i = \pm 1/2$). We used a complete basis set for the entire system at small $N$ and for the largest $N$, complete basis sets for given values of $MS_{Tot}$ ($MS_{Tot}$ is the sum of all the $M_i$). The diagonalization yields a large number of excited states, rapidly increasing with $N$, among which only a limited number are actually excited by tunnelling electrons, as will be shown below. The study of the magnetic structure of ensembles of spin ½ has already been studied with Hamiltonian (1) and (2) for simple chains a long time ago (see e.g. Ref.[39,40,41] or in text books[29,30]). Below we present the ring case in more detail and in particular its ability to represent an infinite chain system. We assume the system temperature to be 0 K, i.e. that the system is initially in its lowest energy state. For finite T, infinite systems are much perturbed by spin wave population in the absence of magnetic anisotropy[42]. We did not try to model these problems and concentrated on finite size systems at 0 K, the limit of which when $N \rightarrow \infty$ is an idealized 1D (anti-)ferromagnet. We can stress that the present work makes use of a complete diagonalization of the Heisenberg Hamiltonian, it thus a priori handles all the possible excitations of the magnetic chains, encompassing in particular the 'usual' spin wave mode as described by des Cloizeaux and Pearsson [39] and the two- and multiple-spinon states discussed by Karbach al[32,33].

In the ferromagnet case, the structure is simple: the ground state corresponds to the maximum possible total spin of the system, $S_{tot} = N/2$, and it is fully degenerated in the present simple system without anisotropy that we study. Actually a small magnetic field ($10^{-6}$ T) was added in the calculation to slightly split the states and thus allow for an easy labelling. The ground state is then the state with $S_{tot} = N/2$ and $MS_{Tot} = N/2$ ($MS_{Tot}$ is the projection of the total spin on the quantization axis). Below, we consider a vanishing temperature and only this ground state is populated; including the other $S_{tot} = N/2$ substates is not thought to modify the results presented below, except for the situations where an explicit polarization of the tunnelling electrons is considered (not discussed here). The total energy per atom of the ground state is independent of the number of atoms and equal to $J/4$ in the ring case. It is equal to $\dfrac{J(N-1)}{4N}$ in the chain case and goes to $J/4$ as $N \rightarrow \infty$ in the open chain case. In the case of a finite ring, the eigenstates of (2) correspond to special solutions of the infinite system and analysis of the symmetry of the eigenfunctions with respect to translation allows to assign a wave number to them, $k$ (see the discussion in Ref.39), so that the diagonalization of (2) directly yields the dispersion relation of the spin excitations in the infinite system.

Figure 1 presents the lowest eigenstates of a ferromagnetic ring with N = 14. The ground state corresponds to $S_{tot} = 7$ and the first simple excited states are $S_{tot} = 6$ states. The known dispersion of 1-dimensional spin waves in an S=1/2 ferromagnet is given by[43] :

$$E(k) = 2J \sin^2(\frac{k\,a}{2})  \qquad (3)$$

One can see in Fig.1 that a set of states with $S_{tot} = 6$ obtained from (2) (actually the lowest one in this symmetry for each $k$) reproduces perfectly the spin wave dispersion; the corresponding results for the spin wave states obtained with N =12 and 10 are also shown in the figure (note that they correspond to $S_{Tot} = 5$ and 4, resp.). We can then conclude that the states of the finite size ring correspond to quantized spin waves in the ferromagnet case or in other words that the ring provides a discretization of the spin wave continuum in the infinite chain system.

The situation is quite different in the anti-ferromagnet case as is well known, due to long-range correlations[44,29]. Figure 2 shows the lowest lying eigenvalues of Hamiltonian (2) for an anti-ferromagnet with N = 14. The ground state corresponds to $S_{tot} = 0$. Again the translation symmetry properties of the eigenfunctions (see the discussion in Ref.39) allow to assign a wave-number, $k$, to each eigenvalue and the first excitation for each $k$ value is found to correspond to $S_{tot} = 1$. The energy of the lowest excited state for several values of N (N = 12, 16 and 18) is also shown together with the expected dispersion law of spin waves in this system[39,44,29]:

$$E(k) = \frac{\pi\,J}{2} \sin(k\,a)  \qquad (4)$$

In the absence of magnetic anisotropy (spin-orbit coupling) there is no gap in the spin wave spectrum at $k = 0$ [29]. The energy of the lowest $S_{tot} = 1$ state exhibits a behaviour as a function of $k$ similar but not identical to that predicted for spin waves (equ. (4)). The difference is particularly visible at large $k$ and decreases as the number of atoms in the ring increases (see in particular the state energy for $k = 1$ in Fig.2). This is due to correlations between the various sites in an infinite chain that are not properly accounted for in a finite system. The states determined for a finite size ring do not correspond exactly to quantized spin waves as for a ferromagnet, though we will see below that information on infinite systems can be extracted from them. The same kind of convergence towards the infinite system can be seen by looking at the energy per atom in the ground state of an anti-ferromagnet chain or ring shown in Fig.3. In contrast to the ferromagnet case, the energy per atom in the ground state or a ring of atoms is not independent of the chain length, though it converges toward the exact value for the infinite chain[39] when the number of atoms, N, increases. Not surprisingly, the convergence of the ground state energy is faster in the ring case ($1/N^2$ behaviour) than in the even open chain case ($1/N$ behaviour) as already discussed e.g. in Ref.[39]. Figure 3 also shows results obtained for open chains with an odd number of atoms. The ground state is a doubly degenerate $S_{tot}$ = 1/2 state and the corresponding energy per atom also converges as $N \rightarrow \infty$ toward the common limit, though in a slower way than with an even number of atoms.

## 2.2 Magnetic excitation by tunnelling electrons

In earlier studies[20,21], we showed how to treat the magnetic excitation of a local spin, carried by a single adsorbate, by tunnelling electrons in a strong coupling approach. A detailed presentation of the strong coupling approach for magnetic excitations can be found in Ref.[21]. For a chain of magnetically coupled atoms (a chain of local spins) and for an exciting tip centred on one of the atoms, the same type of approach can be used: the tunnelling process is very fast, much faster than the effect of the interactions between neighbouring magnetic atoms and so, the tunnelling process can be described without inter-spin interactions and the Heisenberg spin-spin couplings are added before and after tunnelling in the sudden approximation. The present treatment thus parallels our earlier one, the magnetic exchange interaction between neighbouring atoms in the chain playing the same role as the magnetic anisotropy terms in [20,21]. Equivalently, one can say that, in both cases, a magnetic atom is interacting with its surroundings (magnetic anisotropy due to the substrate or neighbouring magnetic atoms in the chain). And in both cases, the influence of the atom surroundings on inelastic tunnelling is taken into account via the sudden approximation.

The states in the basis set for describing the system are labelled $B_j$ and defined as:

$$\left| B_j \right\rangle = \left| M_1, M_2, \ldots, M_{N-1}, M_N \right\rangle \qquad (5)$$

Diagonalization of the Hamiltonian (1) or (2) in this basis yields the set of magnetic states of the system labelled $\phi_n$ with energies $E_n$ and expressed in the $B_j$ basis as:

$$\left| \phi_n \right\rangle = \sum_j C_{nj} \left| B_j \right\rangle \qquad (6)$$

We suppose that the STM tip is above the site 1 of the chain (ring). Tunnelling of an electron through the adsorbate 1 occurs very fast so that we neglect the action of the Heisenberg Hamiltonian during tunnelling. The tunnelling amplitude then depends of the spin coupling between the electron and the spin of atom 1. Both being spins 1/2, only two values are possible for the total spin, $S_T$, of the 'electron+atom 1' system: zero or one. For the present model study, we assume that tunnelling is dominated by only one of the two possible symmetries, as was found in our studies on Mn, Fe and FePc adsorbates (this might not be a general result) and we chose the total spin zero (similarly to the case of CoPc chains studied in [5,21]). The tunnelling amplitude from tip to substrate can then be written as:

$$T_{Tip \rightarrow Sub} = \left| S_T = 0 \right\rangle T^0_{Tip \rightarrow Sub} \left\langle S_T = 0 \right| \tag{7}$$

The initial state of tunnelling with the Heisenberg terms taken into account can be written as a product $\left| i \right\rangle = \left| k_i \sigma_i \right\rangle \left| \phi_i \right\rangle$ (and similarly for the final state $\left| f \right\rangle$) where the first ket refers to the tunnelling electron ($k$ is the electron momentum and $\sigma$ is for spin up or down) and the second term is for the chain of atoms. Using the results of the Hamiltonian diagonalization (6) and Clebsch-Gordan coefficients, these states can be expressed as:

$$\left| i \right\rangle = \left| k_i \sigma_i \right\rangle \left| \phi_i \right\rangle = \sum_m A_{i,m} \left| S_T = 0 \right\rangle \left| m = \left\{ M_2, M_3, \ldots, M_N \right\} \right\rangle \tag{8}$$

and similarly for the final state. In the sudden approximation, the tunnelling probability of an electron associated to a transition from state $\left| init \right\rangle$ to state $\left| final \right\rangle$ of the chain of atoms is given by:

$$P(k_i, \sigma_i, \phi_i \rightarrow k_f, \sigma_f, \phi_f) = P(init \rightarrow final) =$$
$$\left| \left\langle f \right| \left| S_T = 0 \right\rangle T^0_{Tip \rightarrow Sub} \left\langle S_T = 0 \| i \right\rangle \right|^2 = \left| T^0_{Tip \rightarrow Sub} \right|^2 \left| \sum_m A^*_{f,m} A_{i,m} \right|^2 \tag{9}$$

The various possible excitations associated to tunnelling then appear as a global tunnelling, $\left| T^0_{Tip \rightarrow Sub} \right|^2$ that is shared among the various magnetic states of the system. If we further assume that this global tunnelling is independent of the tunnelling energy in the small energy range that is spanned in a magnetic IETS experiment, then the conductance of the system (elastic and inelastic) as a function of the junction voltage, V, for unpolarized electrons can be written as:

$$\frac{dI}{dV} = C_0 \frac{\sum_f \Theta(eV - E_f) \sum_{\sigma_i, \sigma_f} \left| \sum_m A_{i,m} A^*_{f,m} \right|^2}{\sum_f \sum_{\sigma_i, \sigma_f} \left| \sum_m A_{i,m} A^*_{f,m} \right|^2} \tag{10}$$

where $C_0$ is the global conductance, $E_f$ are the various excitation energies of the system and $\Theta$ is the Heavyside function. Note that this conductance corresponds to a given initial state, noted, $\phi_i$; below we will consider a vanishing temperature so that only one state, the ground state has to be considered as the initial state of the tunnelling process. When the tip bias is above all the inelastic magnetic thresholds of the system, the conductance is equal to the global term $C_0$ and one can define the excitation probability by a tunnelling electron when all magnetic channels are open:

$$P(i \rightarrow f) = \frac{\sum\limits_{\sigma_i, \sigma_f} \left| \sum\limits_m A_{i,m} \, A^*_{f,m} \right|^2}{\sum\limits_f \sum\limits_{\sigma_i, \sigma_f} \left| \sum\limits_m A_{i,m} \, A^*_{f,m} \right|^2} \qquad (11)$$

In equation (11), the electron spin in the initial and final states can be different or identical ($\sigma_i = \pm \sigma_f$), corresponding to the existence of excitation processes (final state different from the initial state) associated or not with a spin-flip of the tunnelling electron (see an example for Fe adsorbates in Ref.[20]).

The mean energy lost by a tunnelling electron (transferred to magnetic excitation) when all magnetic channels are open is given by:

$$\Delta E_{mean} = \sum_f (E_f - E_i) \, P(i \rightarrow f) = \frac{\sum\limits_f (E_f - E_i) \sum\limits_{\sigma_i, \sigma_f} \left| \sum\limits_m A_{i,m} \, A^*_{f,m} \right|^2}{\sum\limits_f \sum\limits_{\sigma_i, \sigma_f} \left| \sum\limits_m A_{i,m} \, A^*_{f,m} \right|^2} \qquad (12)$$

All these expressions are for non-polarized electrons and can be easily generalized to the 'polarized case' by changing the sums over $\sigma_i$ and $\sigma_f$ (see e.g. in [22] the application of the polarized strong coupling approach to the case of the Mn/CuN system that has been experimentally studied by IETS with a polarized tip[45]). All the results shown below for ferromagnetic and anti-ferromagnetic cases were obtained in the non-polarized case.

One can stress here the qualitative view of the excitation of a magnetic chain by a tunnelling electron that is carried by the present strong coupling approach. The tunnelling electron only interacts with a given site in the chain during the fast tunnelling process; during tunnelling, the interaction with the electron can flip or not the spin of the active site, all the other sites being spectators. At the end of the collision, the spin flip of a single given site results in the excitation of a whole spectrum of excited magnetic states of the chain. Which states are excited depends on the details of the chain structures (ferro or antiferro, open chain or ring, initial state). Non-spin flip excitations are also possible and are discussed below. The efficiency of the process only depends on the spin coupling ($S_T$ symmetry) during tunnelling and in particular it does not depend on the strength of the magnetic exchange interaction, J.

## 3. Excitation of a ferromagnetic chain

The conductance of a ferromagnetic chain of 12 atoms is presented as a function of the junction bias, $V$, in Figure 4 (the exchange interaction, $J$, is used as an energy unit). Three different cases are presented: a ring of atoms and a linear chain with the STM tip above an end atom and a linear chain with the STM tip above a central atom (here and below, for an even number of atoms in the chain, we will call 'central' the atoms the closest to the geometrical centre, although there is no real central atom; in Fig. 4, with $N = 12$, it is the sixth atom). In all cases, the conductance has been normalized to one for voltages above the magnetic excitation thresholds (i.e. $C_0 = 1$ in equation (10)). Only positive biases are presented, the behaviour for a negative bias is symmetric. The conductance as a function of the tip bias presents a series of steps that are associated with the energy thresholds for the magnetic excitation in the system. This is the essence of the magnetic IETS[1,2,3,4,5,6,7,8] which allows the experimental determination of the energy spectrum of magnetic excitations. When the tip voltage is increased, the conductance exhibits a sharp up-jump when an excitation threshold is met, i.e. when a new inelastic channel is open, allowing more electrons to flow from tip to substrate; the height of the conductance jump yields the importance of this particular excited state in the conductance, i.e. is related to the excitation probability of that particular excited magnetic state by a tunnelling electron. No broadening effect is included in the present calculation so that the steps in the conductance spectrum are infinitely sharp; the finite slope is due to the finite voltage grid in the calculations. The conductance for zero bias is due to elastic tunnelling, it amounts to 0.5 in Fig.4 and the successive steps observed in the tunnelling conductance correspond to the successive magnetic excitation thresholds of the chain (at least those of the magnetic states that are excited by a tunnelling electron). The total contribution from the inelastic steps amounts to 0.5, so that in this model system, at large bias, elastic and inelastic tunnelling are equally probable. It is worth noting that in the $N = 12$ case depicted in Fig.4, the ground state is a state of symmetry $S_{tot} = 6$ and $MS_{Tot} = 6$; the first step in the conductance appears at the first point in the bias grid and corresponds to the excitation inside the $S_{tot}$=6 manifold to the $MS_{Tot}$=5 state. This magnetic transition appears as an inelastic process due to the very small finite B field that has been added to the system to split the ground state.

One can see that, although many states are excited, not all the magnetic excitations of the chain are present, only a small fraction of magnetic states are excited and in particular, the excitation stops above energies larger than twice the exchange interaction, J. One can also notice that exciting a linear chain at one end or at the middle results in different individual excitation probabilities, although the total inelastic probability remains the same, 0.5. So, similarly to the case of individual magnetic atoms, we find for a chain of atoms that inelastic tunnelling is a very large (50%) fraction of the total tunnelling and further more that it is independent of the chain length so that we can conclude that for an infinite chain we would find an inelastic probability of 0.5, too.

In a ferromagnetic chain, the ground state corresponds to all spins in the chain being parallel ($S_{Tot}= N/2$). In the present model system with a very small magnetic field, the ground state is fully polarized along the field axis. The only transition that can be excited by the fast tunnelling of an electron through one of the atoms is then the spin flip of this atom, which in turns corresponds at the end of the collision to exciting states with $S_{Tot} = N/2$ or $N/2 -1$ in the chain. Thus, in this case, inelastic tunnelling is entirely of the spin-flip type. This qualitative view corresponds exactly to the mean energy that is lost by the tunnelling electron. The mean energy loss (Equ.(12)) is exactly equal to $J/2$ for the ring, to $J/2$ for a linear chain with the STM tip above a central atom and to $J/4$ for a linear chain with the STM tip above an end atom. These mean energy losses are independent of the number of atoms in the chain and correspond exactly to the flip of a local spin in the ground state configuration. One can also express the above results in term of spin transfer torque[19]; in the present ferromagnetic model system, for a tunnelling electron with a spin polarisation opposite to that of the chain and an energy higher than the inelastic thresholds, the transferred momentum is equal to 0.5 ħ/electron.

The actual distribution of the excitation among the excited states however depends on the number of atoms in the chain. Figure 5 presents the conductance of a ring of atoms with different atom numbers: 10, 12 and 14. The excitation energies of the different rings are different (see e.g. Fig.1) and so the steps in the conductance vary both in position and height with the number of atoms in the ring. Though, only looking at Fig.5, one can say qualitatively that the three step curves correspond to different discretizations of a single continuous curve, which would be the result for an infinite number of atoms. The same result (not shown here) is found for the excitation of a linear chain (end atom or central atom excitation); increasing the number of atoms in the chain leads to a better representation of the continuous conductance of an infinite system. One can also stress the fact that the ring conductance is very close to the conductance of a linear chain with the STM tip on the central atom (see Fig.4), stressing that the primary excitation is local and thus independent of the chain edges (open or closed). We could try to go further; using a very large number of atoms and smoothing the conductance as a function of bias, we could generate a continuous curve that would represent the excitation of an infinite chain, i.e. the excitation of spin waves. In fact, the derivative of the continuous conductance with respect to the bias yields the energy spectrum of excited spin waves (equivalently, the energy loss spectrum of the tunnelling electron). In the case of a ferromagnetic chain, a much easier method can be used. Comparing Fig.1 and 5, one can see that only the states of the ring corresponding to the spin waves, i.e. to the dispersion curve given by Equ. (3), are actually excited. This is not surprising in view of the qualitative picture of the process and since the finite size calculation reproduces exactly a finite subset of the eigenstates of the infinite system. Furthermore, it appears in our calculation that all the quantized spin waves that appear for a finite ring are equally populated by a tunnelling electron (actually the probability for the states at the two ends of the spectrum are 50% smaller) ; this feature can also be seen in the conductance spectra in Fig.5. As a consequence we can conclude that a tunnelling electron excites a white $k$-spectrum of spin waves in an infinite ferromagnetic chain. From this, knowing the dispersion (Equ.(3)) of spin waves, $E(k)$, we can deduce the energy spectrum $Sp(E)$ of the excited spin waves:

$$Sp(E) = Sp(k)\frac{dk}{dE} = C\frac{1}{\sqrt{E}\sqrt{2J - E}} \qquad (13)$$

The constant $C$ is such that the integral of the spectrum represents half of the total tunnelling current. From (13), we can deduce the conductance $dI/dV$ of an infinite chain as:

$$\frac{dI}{dV} = 0.5\left(1 + \frac{2}{\pi}Arc\sin(\sqrt{\frac{eV}{2J}})\right) \qquad (14)$$

This prediction for the infinite chain has been plotted in Fig.5 together with the results for finite rings, it is seen to nicely represent the expected continuous limit of the step curves obtained for finite rings.

To summarize the above results on spin wave excitation by a tunnelling electron: i) a finite ring of ferromagnetic atoms do represent exactly a piece of an infinite chain of ferromagnetic atoms and excitation of a finite chain can be described in terms of quantized spin waves of the infinite system, ii) a tunnelling electron efficiently excites spin waves in a ferromagnetic chain (50% probability in the present model system), iii) only the usual spin waves are excited and iv) the excited spectrum of spin waves is constant as a function of the wave number. As for excitation of a finite chain, they bear strong resemblances with spin wave excitations, in particular in the case of an STM tip above a central atom of a linear chain.

## 4. Excitation of a chain of anti-ferromagnetic atoms

The magnetic excitation of an anti-ferromagnetic chain of atoms is completely different from that of a ferromagnetic chain both because of the difference in the structure of spin waves (note in particular the different dispersion relations in (1) and (2)) and the fact that a finite size chain of anti-ferromagnetic atoms is only an approximation for a piece of an infinite chain, slowly converging toward the spin wave case for very long chains (see for ex. Fig.3).

Figure 6 presents the conductance of an anti-ferromagnetic chain of 16 atoms as a function of the junction bias (the exchange interaction, $J$, is used as an energy unit). Three different cases are presented: a ring of atoms, a linear chain with the STM tip above an end atom and a linear chain with the STM tip above a central atom (the eighth atom in the chain, actually). The conductance has been normalized to one for voltages above the magnetic excitation thresholds. Only positive biases are presented, the behaviour for negative bias is symmetric. No broadening effect is included. The elastic contribution to the conductance amounts to 0.25, so that, in this system, at large bias, the inelastic tunnelling is overwhelmingly dominating, being three times larger than elastic tunnelling. This inelastic/elastic ratio is independent of the chain length and then it will represents the infinite chain case. Actually, since we chose a model system with the same spin structure as the chains of CoPc molecules studied in [5,21], we find for the model anti-ferromagnetic chains the same inelastic/elastic ratio than found in the chains of CoPc.

Similarly to the ferromagnetic case, we find that many magnetic states are excited, though not all of them. Similarly too, the conductance of the linear chain for an STM tip above a central atom is different from that of a chain with the STM tip above an end atom and very similar to that of a ring of atoms (note though that the excitation energies are different in the chain and ring cases). Figure 7a and b presents the $N$ dependence (for even $N$) of the conductance spectrum in the chain case (end atom excitation) and in the ring case. The energy position and the number of excited magnetic states vary with $N$, but the conductance spectra for the different $N$ resembles very much one to the other, they appear as different step functions, approximations of the continuous curve that would be the limit for an infinite chain.

Several calculations were performed in the case of a linear chain with an odd number of atoms (rings with odd numbers of atoms and a cyclic boundary condition are impossible). Figure 8 presents the corresponding conductivities. In an odd chain, the anti-ferromagnetic ground state is associated to the total spin $S_{Tot} = 1/2$. The conductivity step at very small bias is associated to the transitions inside the ground state manifold; it appears at finite energy in the figure because of the very small magnetic field that is added in the calculations (see section 2.1). Qualitatively, the results exhibit features similar to those for even chains: i) the total inelastic conductivity is large, ii) excitation on end atoms or central atoms are different, iii) excitations on central atoms correspond on the average to higher energies and iv) the conductivities obtained for different N are very similar, corresponding to different discretization of the same continuous curve. As the main difference, the elastic contribution to tunnelling is larger in the odd number case (50 %) than in the even number case (25 %). In the case of finite length chains, the change of spin symmetry thus leads to alternating behaviours of the system for odd and even numbers. This has been very clearly observed experimentally for anti-ferromagnetic Mn chains[3] as well as in a perturbative study on small systems[16]. The experimental results[3] display the same qualitative features as observed here, i.e. a strong magnetic excitation with a ratio inelastic/elastic conductance exhibiting a weak, if any, dependence on the number of atoms in the chains in each parity group.

Figure 9 presents the mean energy loss of a tunnelling electron (Equ.(12)) as a function of the inverse of the number of atoms in the chain. The five studied systems (ring, linear chain with end excitation and linear chain with central excitation, odd and even number of atoms) are presented. In contrast to the ferromagnetic case, the mean energy losses are not independent of $N$, though they converge at large $N$ toward a well-defined value at infinite $N$. This limit is equal to $0.65 \pm 0.01$ J for the end excitation of an even linear chain. The ring and the linear chain with central excitation and even $N$ converge toward the same limit (or very close limits), of the order of $0.89 \pm 0.01$ J. These energy losses are significantly larger than those found in the ferromagnetic case (0.25 and 0.5 J, resp.). The mean energy losses in the odd $N$ case also seem to converge toward a well defined value around 0.43 J for the end excitation and 0.6 J for the central excitation. Consistently with a comparison between fig. 6 and 8, the mean energy transferred from the tunnelling electron to the magnetic system is smaller for an odd $N$ chain than for an even $N$ chain. Actually, simply rescaling the relative weight of elastic and inelastic contributions to tunnelling for odd and even numbers of atoms (not shown here) makes the conductivity in the two cases very similar (again as different discretizations of the same continuous curve). In the same way, multiplying the mean energy losses for odd numbers of atoms by 1.5 makes them quite consistent with the results for the even chains.

In an anti-ferromagnetic chain, the ground state wave-function cannot be written simply (see discussion in various textbooks e.g.[44,29]). For an even number of atoms, it is a $S_{tot}$ = 0 state. The two configurations formed by alternating spins up and down are not sufficient to describe the ground state of the system, since they are coupled with any other configuration differing by the spin flip of two neighbours. As a consequence, the system ground state is described by a superposition of configurations, and its energy and its wave function converge slowly with the number of atoms included, i.e. the number of configurations, as can be seen in Fig. 2 and 3. For a given site along the chain, the various configurations correspond to either a spin up or a spin down and consequently, we cannot have the simple 'local spin flip' picture that was found in the ferromagnetic case. Let us consider an electron with a spin up incident on the ground state of a chain at a given site; in the fast tunnelling stage, it only couples with the configurations having a spin down at the active site to form the tunnelling symmetry $S_T$ =0; thus only these configurations will be present during tunnelling and will contribute to the population of final states. At the end of the collision, the emitted electron can leave with a spin down leading to a series of spin-flip excitations. But it can also leave with a spin up; however, since the configuration with a spin down at the active site is not an eigenstate of the system, this will lead to another series of excitations of non-spin-flip kind. This second excitation series is a direct consequence of the projection of the initial wave-function on the tunnelling symmetry which rejects half of the spin configurations. Qualitatively, we can say that there are two kinds of excitation: a 'usual' spin flip kind and a non-spin-flip kind induced by the correlation between the various sites. This is illustrated in Fig.10 which shows the spin-flip and non-spin-flip contributions to the conductance in the case of a linear chain of 16 atoms (excitation on an end atom). Both contributions are quite important and they exhibit steps at the same energies. Actually, since excitation by an electron can only populate states with $S_{Tot}$ = 1, the spin-flip and non-spin-flip contributions correspond to the excitation of the substates with $MS_{Tot} = \pm 1$ and $MS_{Tot} = 0$ of the $S_{Tot} = 1$ manifolds, respectively. Furthermore, it appears that the probability for excitation to the two degenerated $MS_{Tot} = \pm 1$ sub-states is twice that for the $MS_{Tot} = 0$ sub-state (for polarized and non-polarized electrons). All these features on spin-flip and non-spin-flip transitions are found in all the systems (chains and ring) with even $N$ studied here. Expressed in terms of spin transfer torque, this means that in the present anti-ferromagnetic system, for electron energy above the inelastic thresholds, the transferred momentum from a polarized tunnelling electron is equal to 0.5 ħ/electron.

The distribution of the excitation among the excited states in the case of a ring of anti-ferromagnetic atoms is not as simple as for the ferromagnetic case. Indeed, the first $S_{Tot} = 1$ set of states, the 'usual spin waves', are not the only states that are excited. This can be seen in Fig. 7b where inelastic steps are found beyond $\pi J/2$, the maximum energy of the first spin wave mode (see fig.2). Information about the spectrum of excited spin waves can be obtained from the conductance in Fig. 6 and 7. Convolution of the conductance as a function of the voltage with a broad enough Gaussian function yields a quasi continuous curve, the derivative of which yields the energy loss spectrum of a tunnelling electron. It corresponds to the spectrum of the energy transferred from a tunnelling electron with an initial energy larger than all the inelastic magnetic thresholds to the magnetic excitations. In other words, it corresponds to the energy spectrum of all the spin waves that are created by a tunnelling electron. Such derivatives are shown in Fig. 11a and b for the linear chain case (end atom excitation) and for the ring case (the central excitation of the linear chain is very close to the ring case). In Fig. 11, the same Gaussian broadening (width equal to *J/3*) has been used for all the chain lengths. For the smallest $N$, remnants of the discrete structure are still present and disappear slowly as $N$ is increased. The convergence is not very fast, as indicated by the magnitude of the convolution width that has to be used. Note that in the present model system, the integral of the energy loss spectrum corresponds to the inelastic probability (0.75) and it should be completed by an elastic peak, i.e. by a delta-function at zero energy loss of integral 0.25. In the case of end excitation of a linear chain, the energy loss spectrum presents a single broad maximum centred slightly below *J*. In the ring case, the maximum around 1.5 *J* is due to the maximum in the *E(k)* dispersion function of the lowest spin wave mode (located at $\pi J/2$, see Fig.2), the structure at low energy comes from the large *k* excitations of the lowest spin wave mode (see below) and the part of the spectrum at high energy corresponds to the excitation of higher spin wave modes. Part of the high energy tail is also due to the Gaussian broadening that has been applied.

Further information about the spin wave spectrum in the ring case can be obtained by analysing the excitation in terms of $k$-states. The various steps in the conductance are associated with various $S_{Tot}$=1 states in the energy spectrum of the ring (Fig.2) and we can extract from it a $k$-spectrum of the excited spin waves. The calculation with a finite ring yields a discretization of the $k$-spectrum; using Stieljes derivative procedure, we can use it to get points in the properly normalized continuous $k$-spectra of the different excited spin wave modes, $Sp_j(k)$ ($j$ is the index of the spin wave mode, see below). Figure 12 presents our results for $Sp_j(k)$ obtained for different N values (N = 12-18). Different spin wave modes, i.e. different states with the same $k$, are excited and we defined the corresponding spectra $Sp_j(k)$ (index $j$) in the following way: for each $k$-value, the state with the highest excitation probability was assigned to the first mode, the second highest probability to the second mode and so on. We chose this somewhat arbitrary procedure rather than the adiabatic definition ($j$-labelling according to the energy order for each $k$) because it leads to more continuous spectra $Sp_j(k)$. Only the three strongest modes were analyzed along these lines. The lowest energy mode, corresponding to the 'usual spin wave', concentrates most of the excitation probability. The three modes that were analyzed appear to be maximum for the largest $k$ values close to the edge of the Brillouin zone ($k = \pi/a$). The dominance of high $k$-values in the $Sp_j(k)$ spectra increases with the index $j$. The results obtained with the various $N$ are close one to the other for the first mode, though one cannot claim convergence of the results for the highest modes with the limited $N$ used here; in particular, the relative weight of the first mode slightly decreases when $N$ increases. Figure 13 illustrates the characteristics of the spin wave modes that are excited by tunnelling electrons. It presents for $N = 18$, the whole set of excited states of the ring case below an energy around $3J$ and highlights the dispersion of the modes that are actually excited. The lowest mode, the one dominating the excitation, is the usual spin wave mode, and the next two are associated with significantly more high lying energies. Comparison of the dispersion curves obtained for N = 16 and 18 for the second and third modes reveals a poor convergence of the weakly excited modes, so that we cannot fully characterize the excited modes with the highest energies. Note that the change in the dispersion function of the second and third mode between low and high $k$-regions seen in Fig.13 might be due to the present somewhat arbitrary definition of the modes. In fact, for large k, these higher modes fall in the range of the two-spinon states analysed by Karbach et al[32,33] in the context of neutron scattering; the continuous character of the two-spinon states most probably accounts for the poor convergence observed in Fig.13: in the present work, the higher mode states being simply a discrete representation of the two-spinon continuum. Some other states, which are excited by tunnelling electrons, are out of the two-spinon range (at small k and/or higher energy) and should then be attributed to higher multiple-spinon states; the corresponding excitation probability is small.

In contrast, the summed $k$-spectrum obtained by summing the first three $Sp_j(k)$ spectra ($j = 1$, 2 and 3) converges much more rapidly. It is shown in Figure 14 and the discrete points in the spectrum obtained with different $N$ numbers appear to fall practically on the same continuous curve, which, then, represents the summed $k$-spectrum of all the spin waves excited by a tunnelling electron (note that Fig.14 uses a linear scale whereas Fig.12 uses a logarithmic scale). Assuming that the excitation is dominated by the two-spinon continuum, one can also say that the $k$-spectrum in Fig.14 corresponds to the distribution inside the two-spinon continuum. The present results can be linked with the earlier results for neutron excitation[32,33] which also reported on an excitation process dominated by the usual spin wave and by the states with a large k.

## 5. Concluding summary

We reported on a theoretical study of magnetic excitations induced in a chain of atoms (ferromagnetic or anti-ferromagnetic) by a tunnelling electron. The studied system is a model one, though it bears some resemblances with the CoPc molecular chains studied in Ref.[5] . The analysis of the results for finite size chains and rings of atoms as a function of $N$, the total number of atoms, allowed us to extract conclusions on the excitation of spin waves by tunnelling electrons in infinite model 1D-systems.

As a striking general result, the excitation of chains (or rings) of different lengths are very similar. Actually, the mean energy lost by a tunnelling electron, the conductance spectrum as a function of the junction voltage and the spectrum of energy loss by a tunnelling electron are very close for different chain lengths. However, the excitation depends on the position of the exciting STM tip along the chain, excitation around the chain centre being very close to the excitation of a ring of atoms. In fact, it is possible to consider that the states in a finite size chain (mainly in a ring) correspond to a quantization of the spin waves of the infinite system induced by confinement in the finite size system. The excitation spectrum for a finite size system can be seen as yielding discrete points in the excitation spectrum of the infinite system; this link is exact in the case of a ferromagnetic ring and only approximate in the case of an anti-ferromagnetic ring.

The excitation probability of spin waves in the chains is very high. We chose a model system, inspired from the system studied in [5], for which the individual excitation probability is very high. This property survives in long finite size chains. The excitation probability for a tunnelling electron energy above the inelastic threshold is equal to 0.5 (ferromagnetic case) and 0.75 (anti-ferromagnetic case) independently of the chain or ring length. We can then conclude on the very large excitation probability of spin waves in an infinite system. The transferred spin momentum is found equal to 0.5 ħ/tunnelling electron in our model system. This result can be linked to the experimental result by Balashov et al[35] on inelastic electron-magnon interaction which also showed a very large spin transfer (of the order of 0.5 ħ/electron) in the case of polarized electrons tunnelling into a ferromagnetic electrode. However, the link can only be qualitative, the experimental system being different from our model system; though, it confirms that tunnelling electrons can be very efficient in exciting spin waves and transferring spin momentum.

The results for ferromagnetic and anti-ferromagnetic chains of atoms are quite different, reflecting the differences between the structures generated by these two types of magnetic coupling.

For *ferromagnetic chains*:

-A finite ring of atoms is equivalent to a piece of an infinite chain. Excitation of a finite size system can be seen as spin waves of the infinite system that are quantized by confinement in the finite size chain.

-Only the lowest spin wave mode is excited (corresponding to a total spin lower than that of the ground state by 1 ħ).

-Excitations of the different $k$-states of the spin wave are equally probable (white $k$-spectrum).

-The picture of the excitation scheme is simple: all excitations are associated with the spin flip of the atom through which the electron is tunnelling. This local spin flip then propagates along the chain and ends up in the excitation of a white $k$-spectrum of spin waves.

For *anti-ferromagnetic chains*:

-A finite ring of atoms is not a piece of the infinite system. However, when the number of atoms is increased, the results obtained with different finite sizes resemble more and more to each other and thus to those of the idealized infinite system.

-The energy transferred from the electron to the magnetic system is larger than in the ferromagnetic case in the present study, in which the spin transfer is modelled in the same way in the two cases.

-Several spin wave modes are excited by tunnelling electrons. In the present studies with a ring with an even number of atoms, the ground state is a singlet state and the excited spin waves correspond to triplet states. The dominant mode is the lowest one, corresponding to the 'usual' spin wave mode. Excitation of all the modes is dominated by large $k$, at the edge of the Brillouin zone.

-No single image of the spin wave excitation appears: part of the excitation is associated to a local spin-flip that propagates along the chain but another part is mediated by correlation between the various sites. This second excitation process is directly linked to the entanglement of the anti-ferromagnet 1D chain.

**Acknowledgments:**

The calculations were partly performed with the GMPCS high performance computer facilities of the LUMAT Fédération of Orsay laboratories.

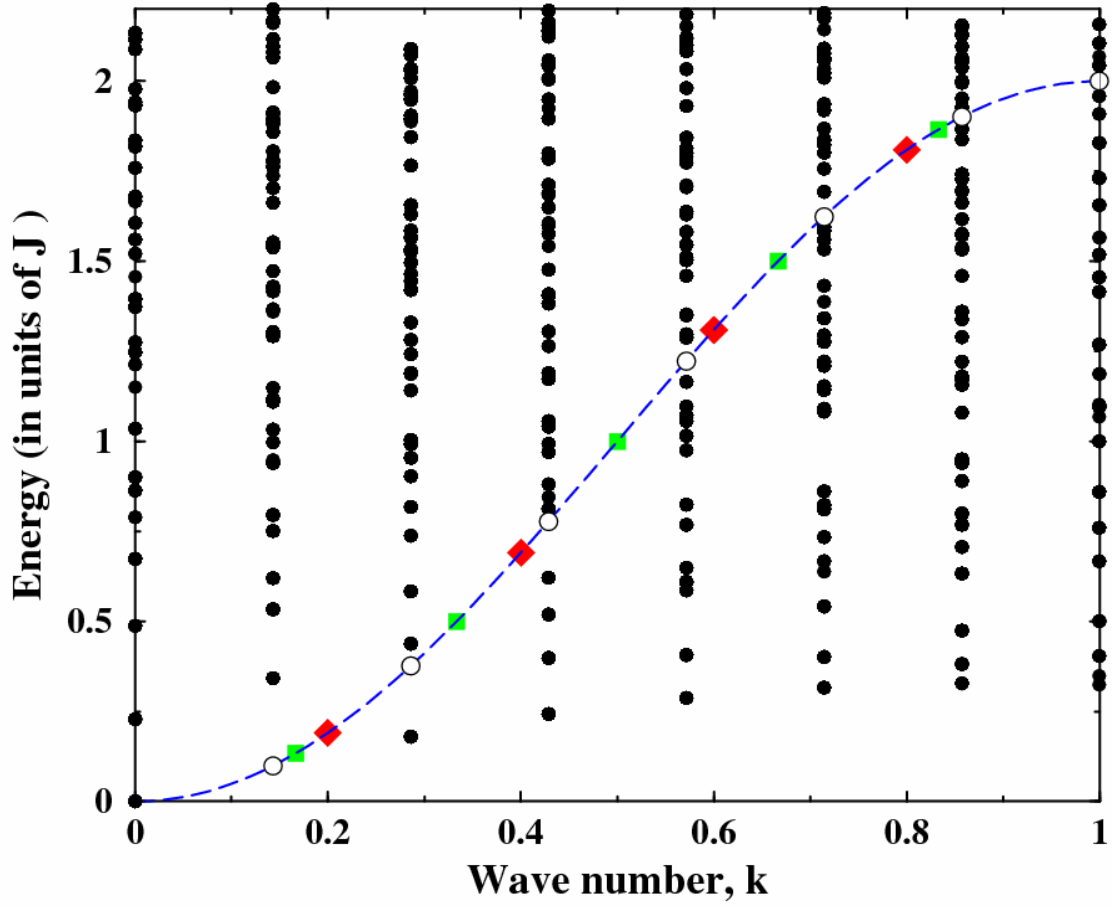

**Figure 1:** Energy of the magnetic levels of a ring of ferromagnetic atoms formed by 14 spins S=1/2. The energy relative to the ground state energy is given in units of the magnetic exchange coupling, *J*, as a function of the wave number in units of π/a (a is the lattice spacing). All the lowest states coming from the diagonalization are shown as full black circles; the states corresponding to quantized spin waves are shown as open blue circles. Results for the quantized spin waves obtained for N = 12 (full green squares) and N = 10 (full red diamonds) are also shown. Dashed blue line: energy dispersion of spin waves in an infinite chain.

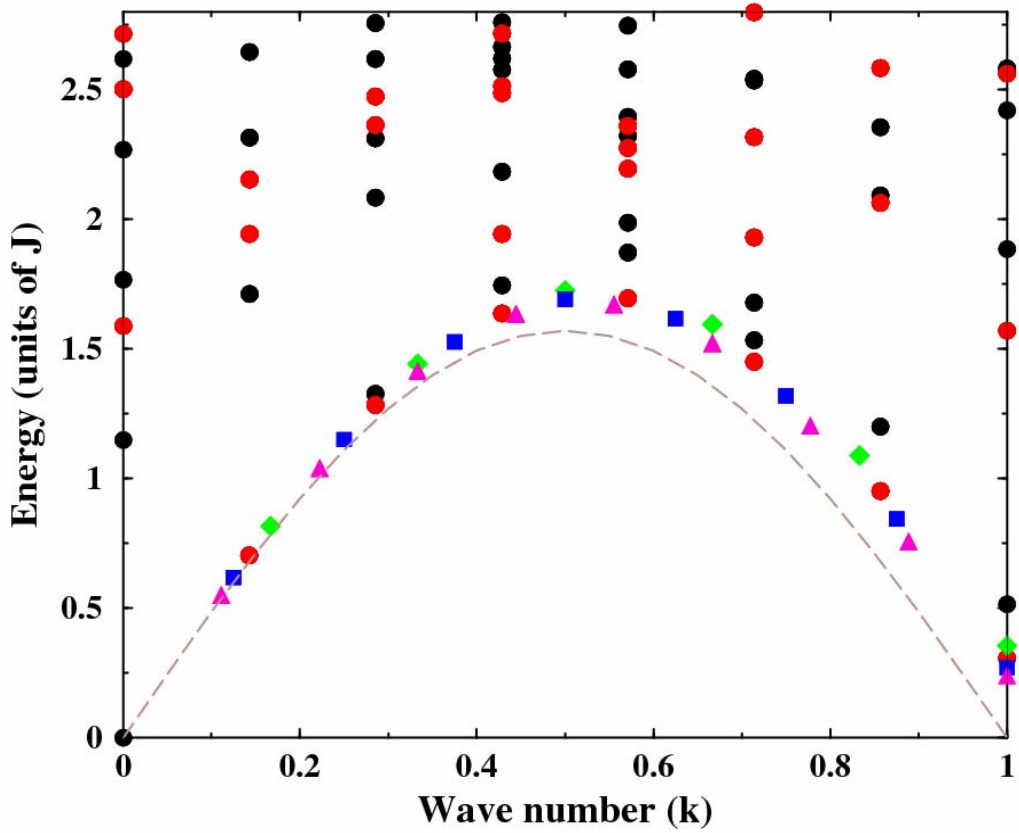

Figure 2: Energy of the magnetic levels of a ring of anti-ferromagnetic atoms formed by 14 spins S=1/2. The energy relative to the ground state energy is given in units of the magnetic exchange coupling, $J$, as a function of the wave number in units of $\pi/a$ (a is the lattice spacing). Full red circles: states with $S_{Tot}$ =1. Full black circles: all other $S_{Tot}$ states for N =14. The lowest $S_{Tot}$ =1 states correspond to the spin wave. The results for the lowest $S_{Tot}$ =1 states obtained with various N are also shown: green diamonds (N=12), blue square (N=16) and magenta triangles (N=18). The dispersion law expected for spin waves in an infinite system is shown by the brown dashed line.

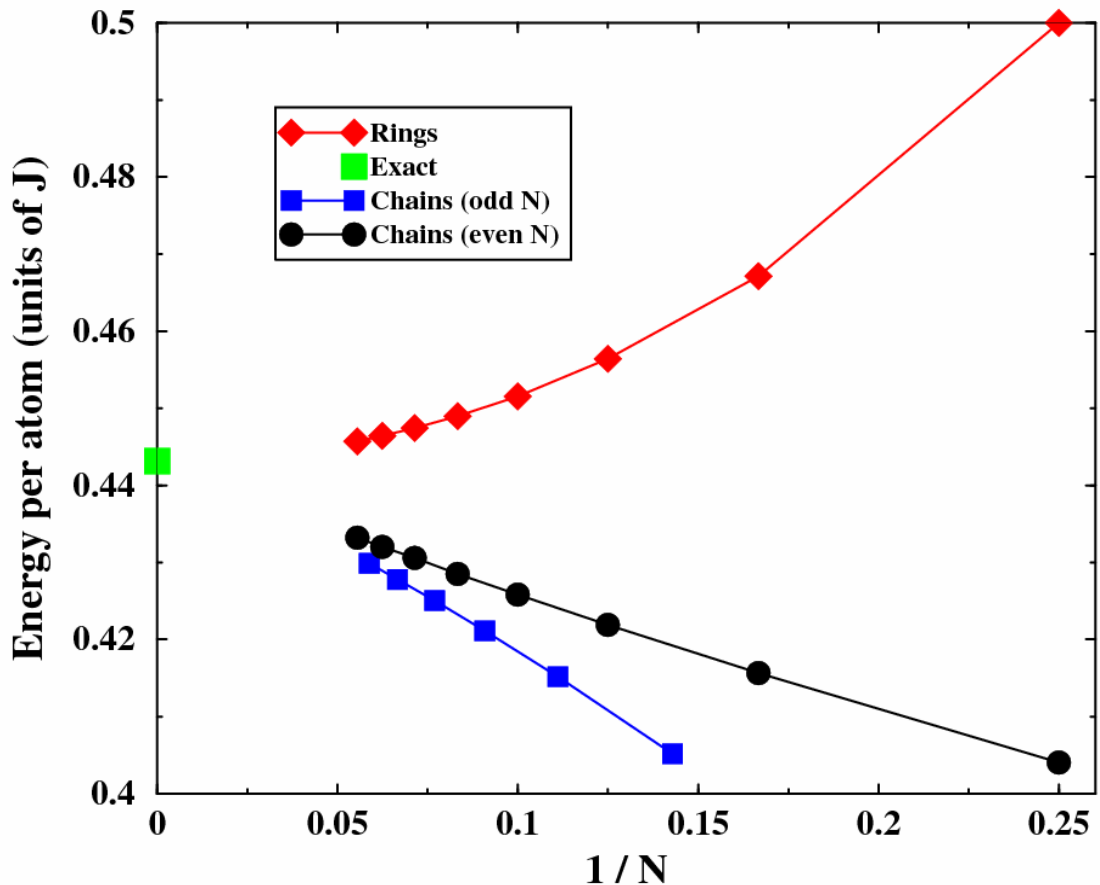

Figure 3: Energy per atom of the ground state of a set of anti-ferromagnetic atoms as a function of the inverse of the atom number, *N*. The energy is shown in units of the magnetic exchange energy, *J*, for the three studied geometries: a ring with an even number of atoms (red diamonds), an open linear chain with an even number of atoms (black circles) and an open linear chain with an odd number of atoms (blue squares). The exact result from des Cloizeaux and Pearsson[39] is also shown for N→∞ (large green square).

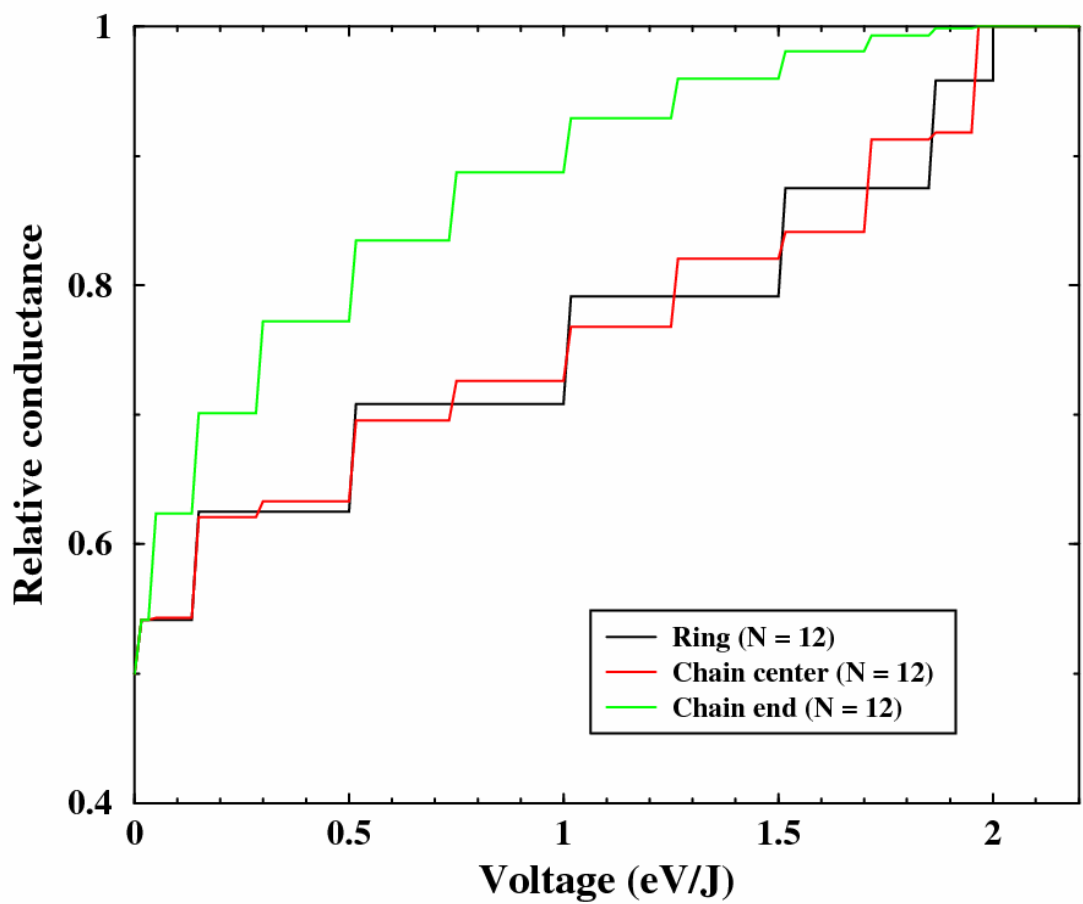

**Figure 4:** Relative conductance of a chain of 12 ferromagnetic atoms as a function of the tip bias (in units of *J*). The conductance has been normalized to one above the inelastic magnetic thresholds. Three cases are presented: a ring of 12 atoms (full black line), a linear chain of 12 atoms with the STM tip above an end atom of the chain (green full line) and a linear chain of 12 atoms with the STM tip above the sixth ('center') atom of the chain (red full line).

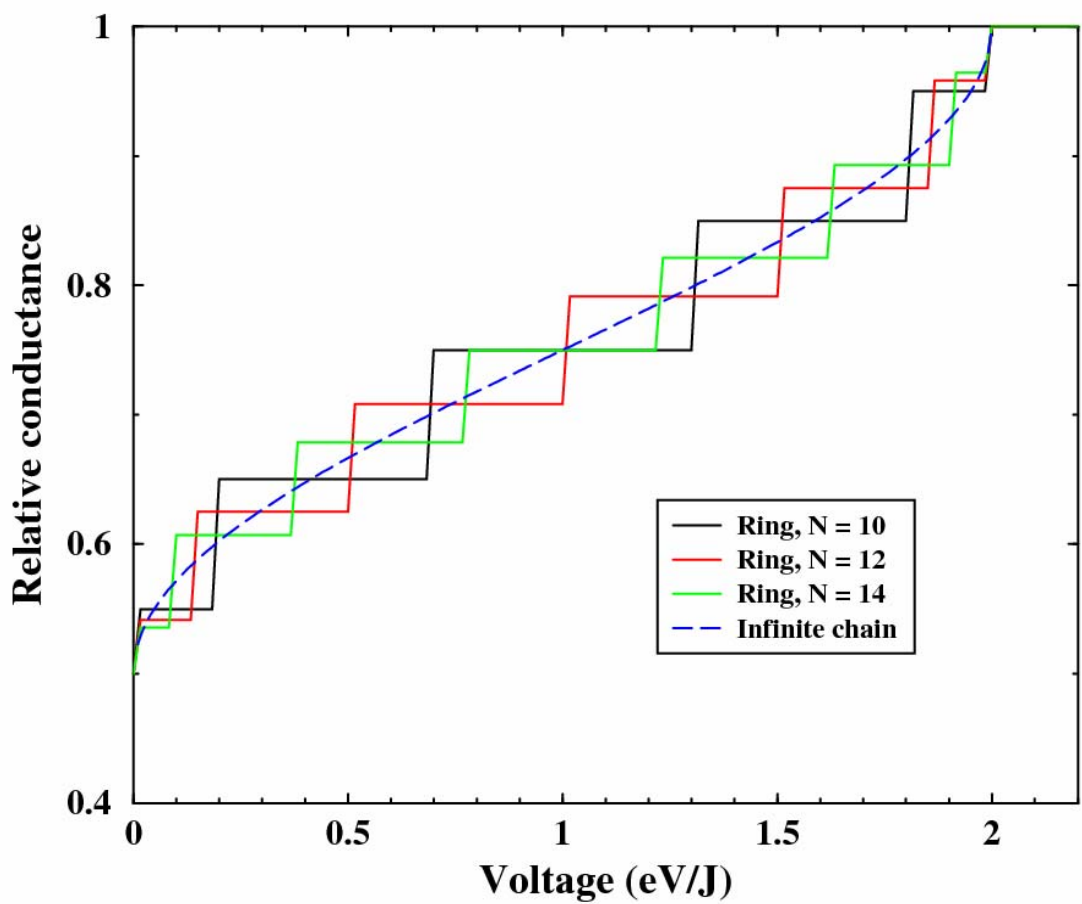

**Figure 5:** Relative conductance of a ring of ferromagnetic atoms as a function of the tip bias (in units of $J$). The conductance has been normalized to one above the inelastic magnetic thresholds. Three different numbers of magnetic atoms are presented: N = 10 (black curve), N = 12 (red curve) and N = 14 (green curve). The dashed blue line corresponds to the conductance of an infinite chain of atoms (see text).

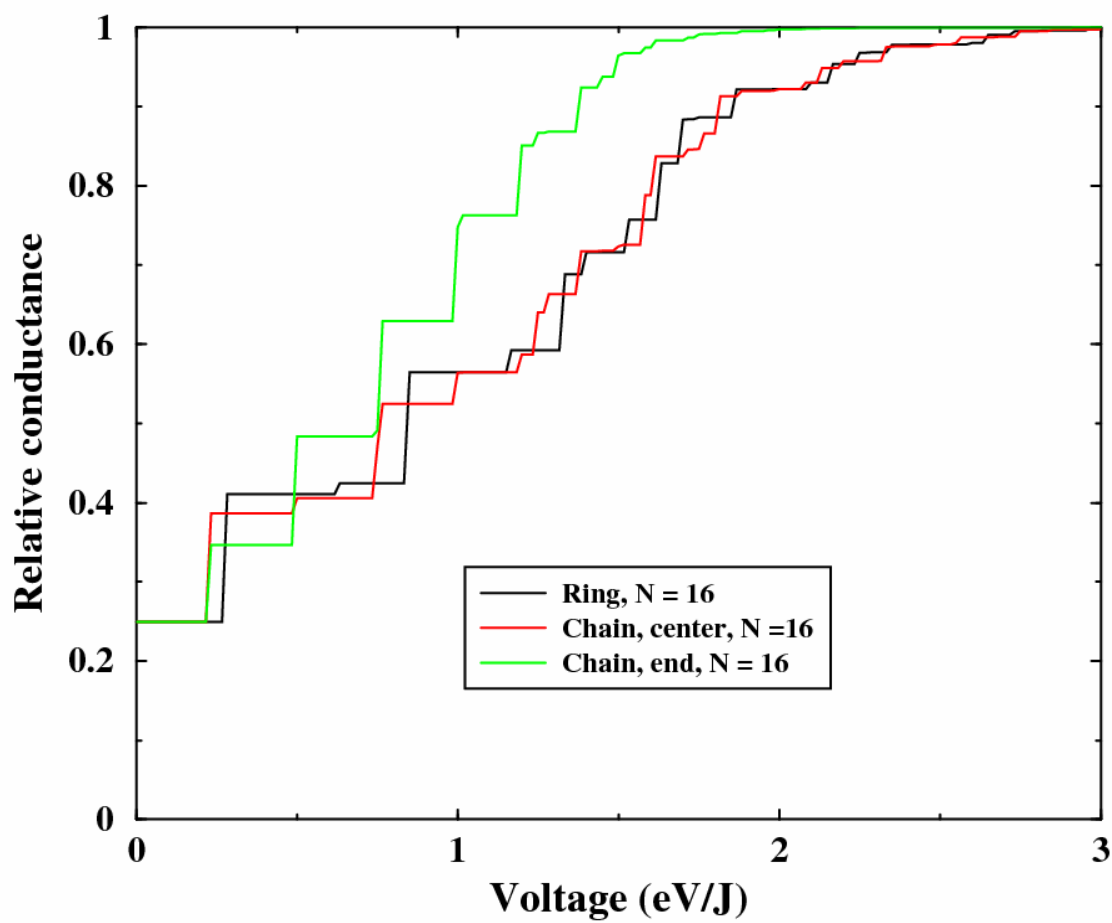

**Figure 6:** Relative conductance of a chain of 16 anti-ferromagnetic atoms as a function of the tip bias (in units of J). The conductance has been normalized to one above the inelastic magnetic thresholds. Three cases are presented: a ring of 16 atoms (black green line), a linear chain of 16 atoms with the STM tip above an end atom of the chain (green full line) and a linear chain of 16 atoms with the STM tip above the eighth ('center') atom of the chain (red full line).

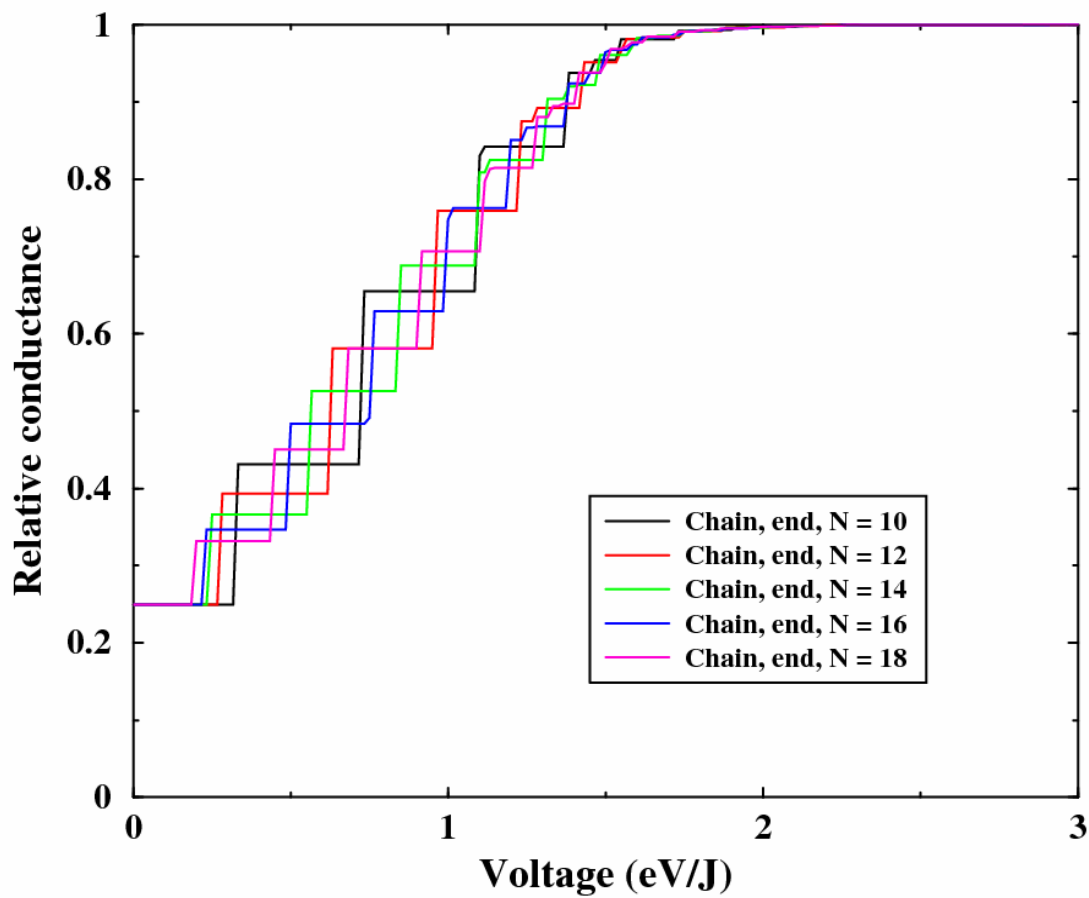

**Figure 7a:** Relative conductance of a linear chain of anti-ferromagnetic atoms as a function of the tip bias (in units of J). The exciting STM is above an end atom of the chain. The conductance has been normalized to one above the inelastic magnetic thresholds. Five different numbers of magnetic atoms are presented: N = 10 (black curve), N = 12 (red curve), N = 14 (green curve), N = 16 (blue curve) and N = 18 (magenta curve).

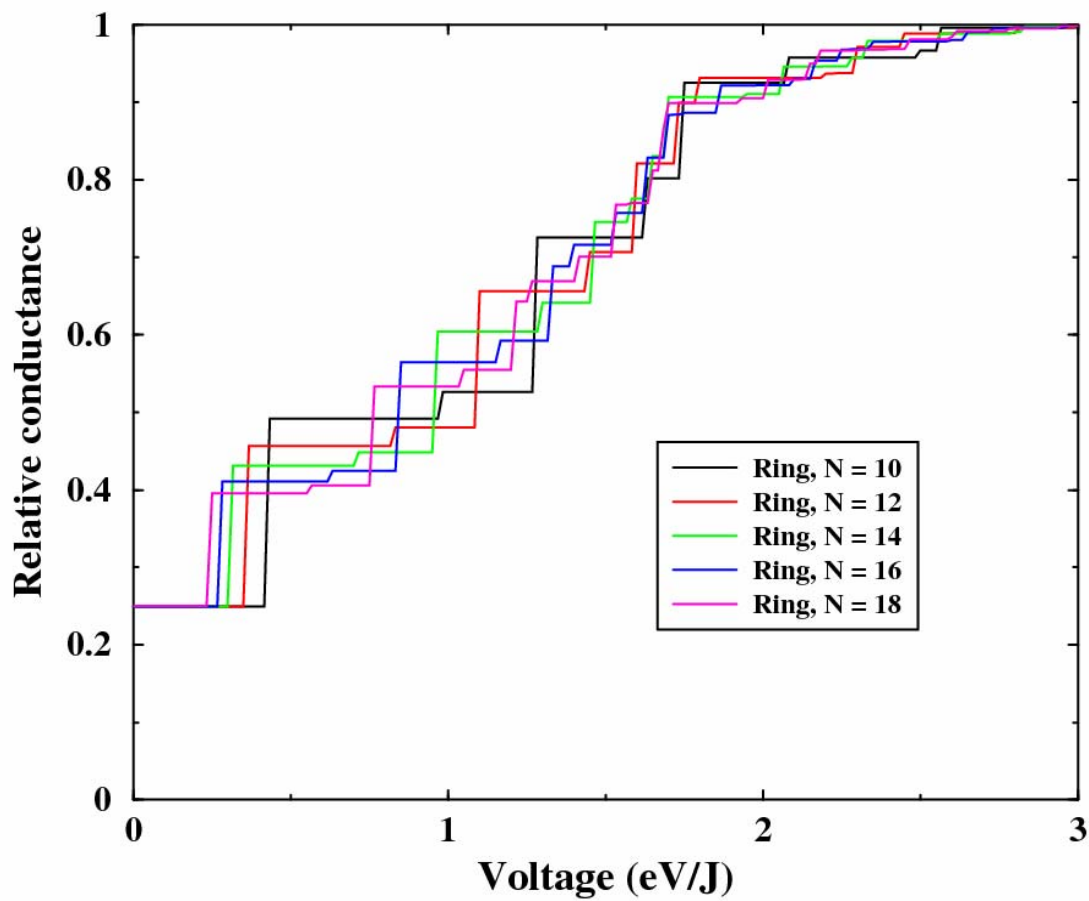

**Figure 7b:** Relative conductance of a ring of anti-ferromagnetic atoms as a function of the tip bias (in units of $J$). The conductance has been normalized to one above the inelastic magnetic thresholds. Five different numbers of magnetic atoms are presented: N = 10 (black curve), N = 12 (red curve), N = 14 (green curve), N = 16 (blue curve) and N = 18 (magenta curve).

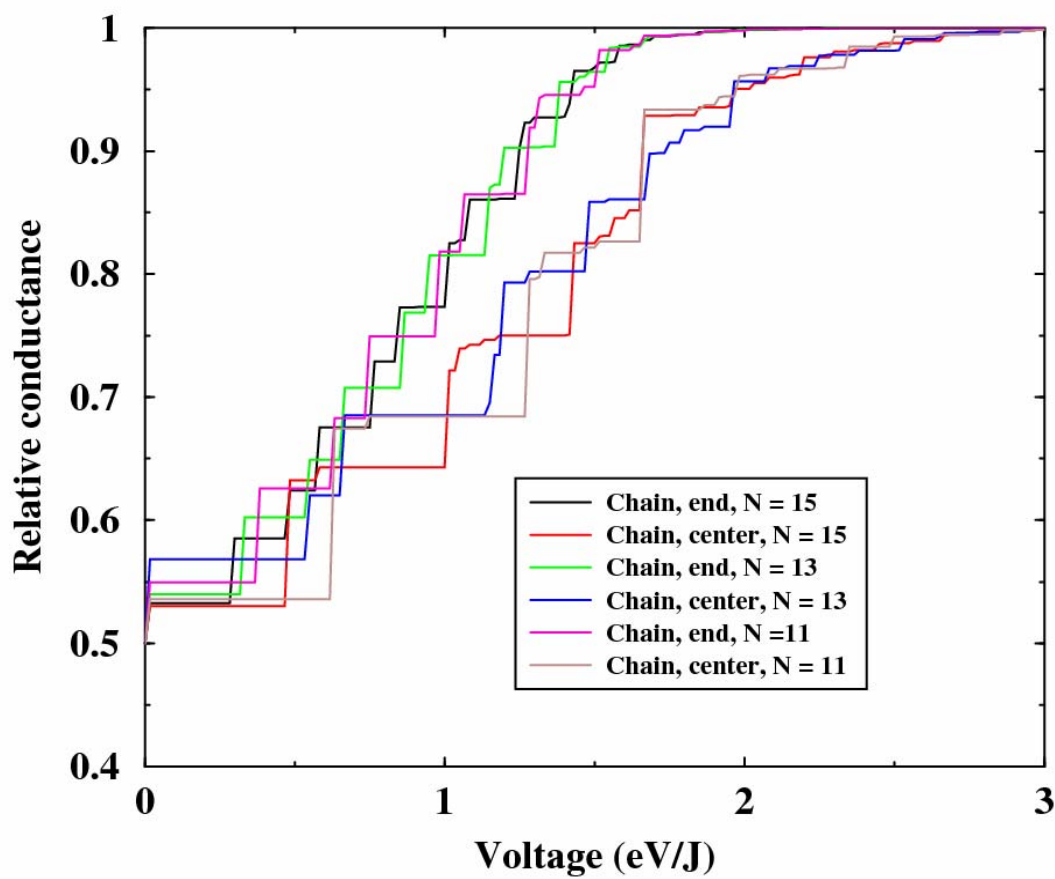

**Figure 8:** Relative conductance of an open linear chain of anti-ferromagnetic atoms with an odd number of atoms as a function of the tip bias (in units of *J*). The conductance has been normalized to one above the inelastic magnetic thresholds. Four different situations are presented depending on the number of atoms in the chain (*N* = 11, 13 and 15) and the position of the STM in the chain (above the central or above an end atom): N = 15, end atom (black curve), N = 15, central atom (red curve), N = 13, end atom (green curve), N = 13, central atom (blue curve), N = 11, end atom (magenta curve) and N = 11, central atom (brown curve).

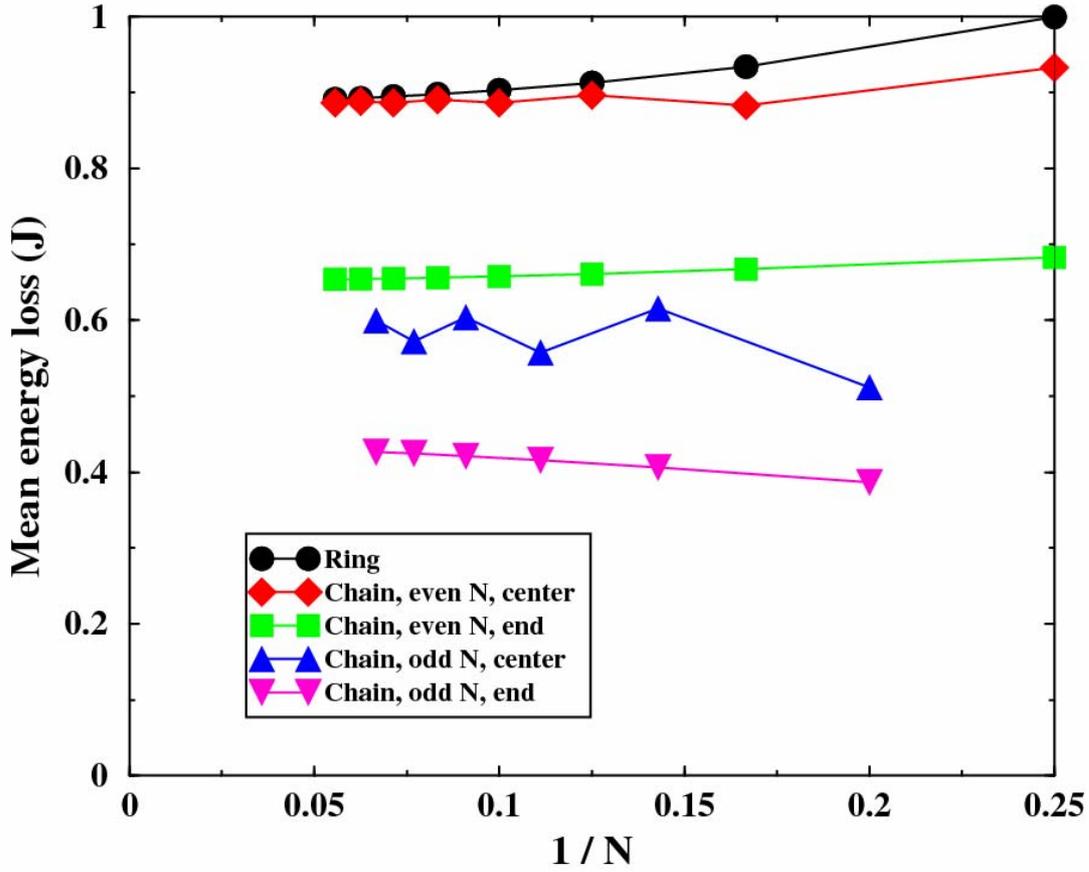

**Figure 9:** Mean energy loss of an electron tunnelling through an atom in an anti-ferromagnetic chain as a function of 1/N, the inverse of the number of atoms in the chain. The mean energy loss is expressed in units of *J*, the magnetic exchange interaction. Full black circles: ring of atoms. Full red diamonds: linear chain with an even number of atoms and the STM tip above a 'center' atom. Full green squares: linear chain with an even number of atoms and the STM tip above an end atom. Full blue up triangles: linear chain with an odd number of atoms and the STM tip above the central atom. Full magenta down triangles: linear chain with an odd number of atoms and the STM tip above an end atom.

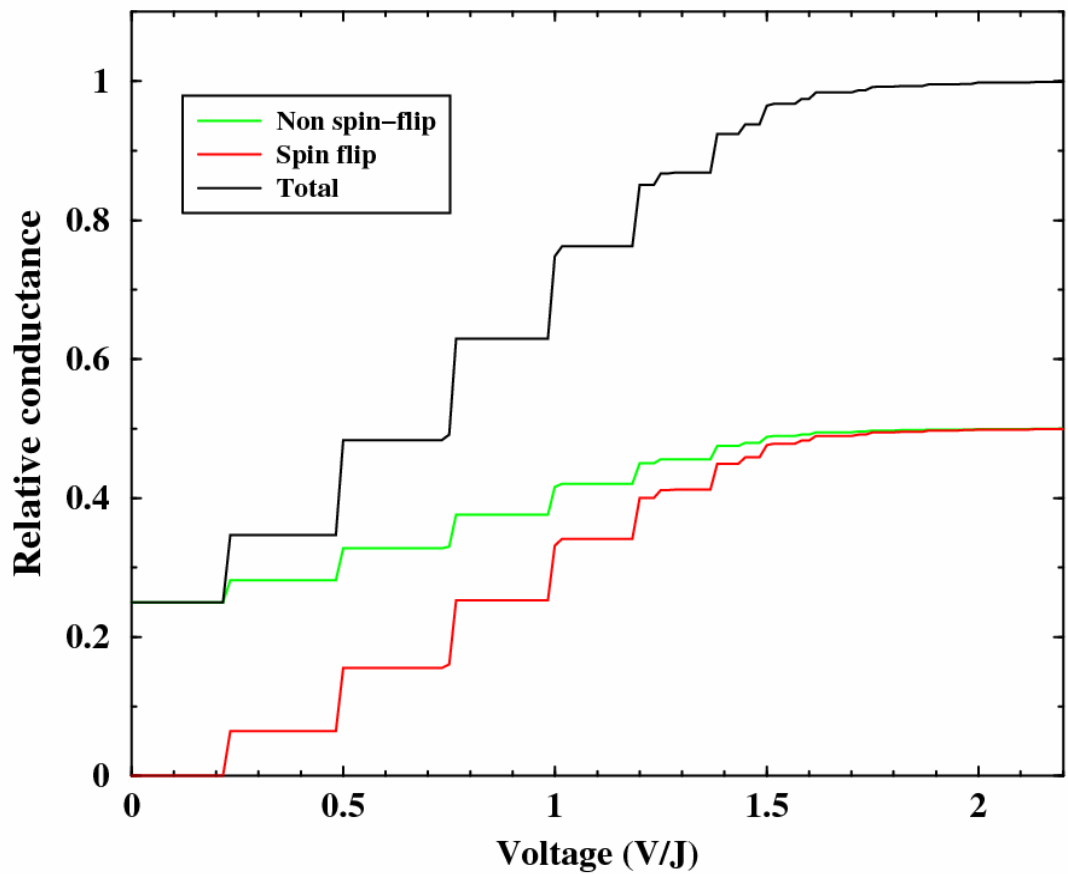

**Figure 10:** Relative conductance of a linear chain of N = 16 anti-ferromagnetic atoms as a function of the tip bias (in units of *J*). The exciting STM is above an end atom of the chain. The conductance has been normalized to one above the inelastic magnetic thresholds. Green line: conductance without spin-flip of the tunnelling electron. Red line: conductance with spin-flip of the tunnelling electron. Black line: total conductance.

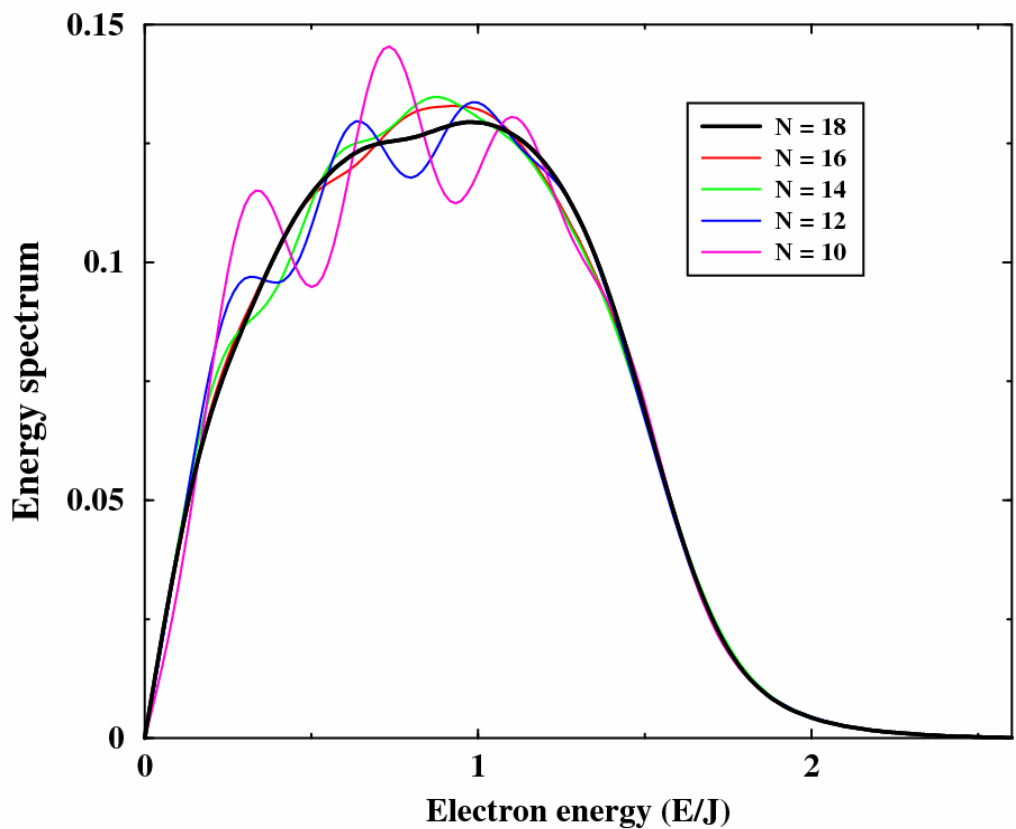

**Figure 11a**: Energy loss spectrum of an electron tunnelling through an end atom of a linear chain of anti-ferromagnetic atoms. The energy of the tunnelling electron with respect to the Fermi level is expressed in units of *J*, the magnetic exchange energy. Five different chain lengths are presented: N = 10, 12, 14, 16 and 18. All the conductance spectra have been convoluted with the same Gaussian of width equal to J/3.

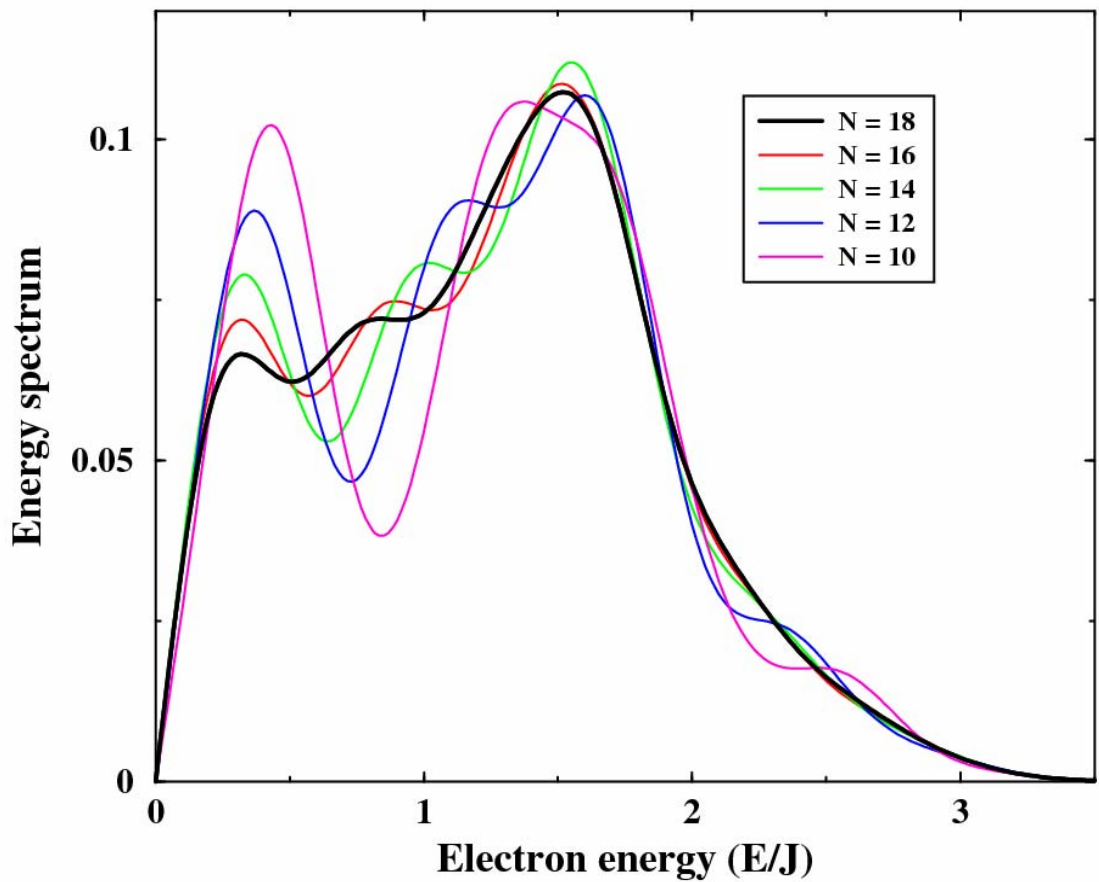

**Figure 11b**: Energy loss spectrum of an electron tunnelling through an atom of a ring of anti-ferromagnetic atoms. The energy of the tunnelling electron with respect to the Fermi level is expressed in units of J, the magnetic exchange energy. Five different chain lengths are presented: N = 10, 12, 14, 16 and 18. All the conductance spectra have been convoluted with the same Gaussian of width equal to J/3.

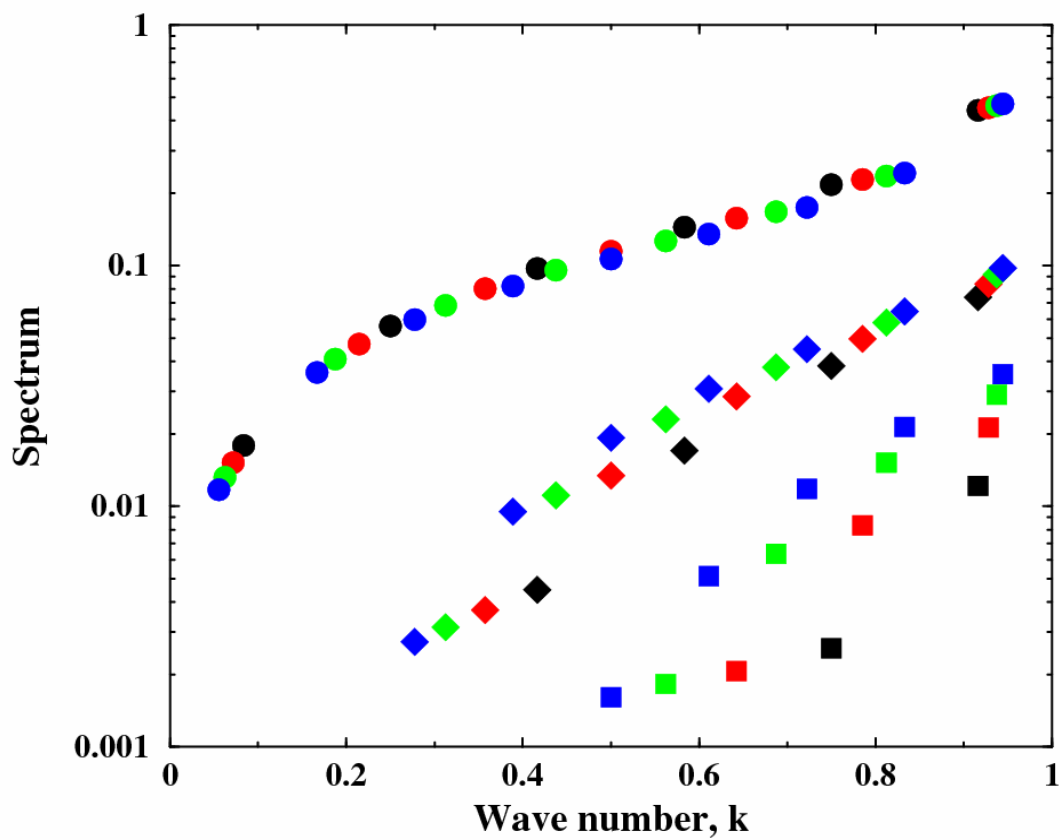

**Figure 12:** Spectrum of the spin waves excited in a ring of anti-ferromagnetic atoms as a function of their wave number. The wave number is expressed in units of π/a where a is the lattice spacing. Three spin wave modes are defined (see text for definition). Coloured circles: first mode, coloured diamonds: second mode and coloured squares: third mode. The results are presented for different numbers of atoms in the ring: N = 12 (black), N = 14 (red), N = 16 (green) and N = 18 (blue).

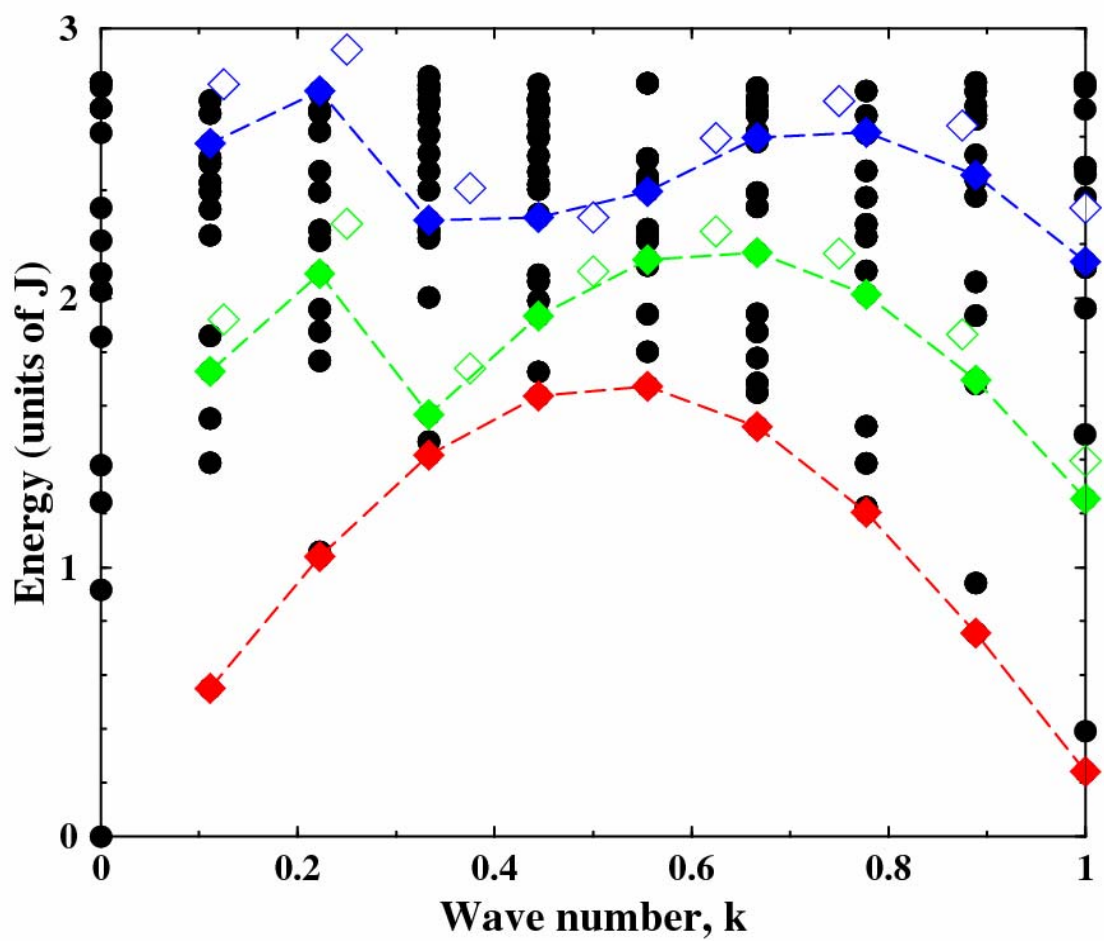

**Figure 13:** Dispersion of the spin waves excited in a ring of N = 18 anti-ferromagnetic atoms. The energy is expressed in units of *J*, the magnetic exchange interaction and the wave number in units of π/a, where a is the lattice spacing. Black circles: all the lowest magnetic states in the ring. Coloured diamonds with dashed lines: states that are significantly excited by a tunnelling electron. Red: first spin wave mode; green: second mode; blue: third mode. The open blue and green diamonds show the dispersion of the second and third mode for a ring of N = 16 atoms.

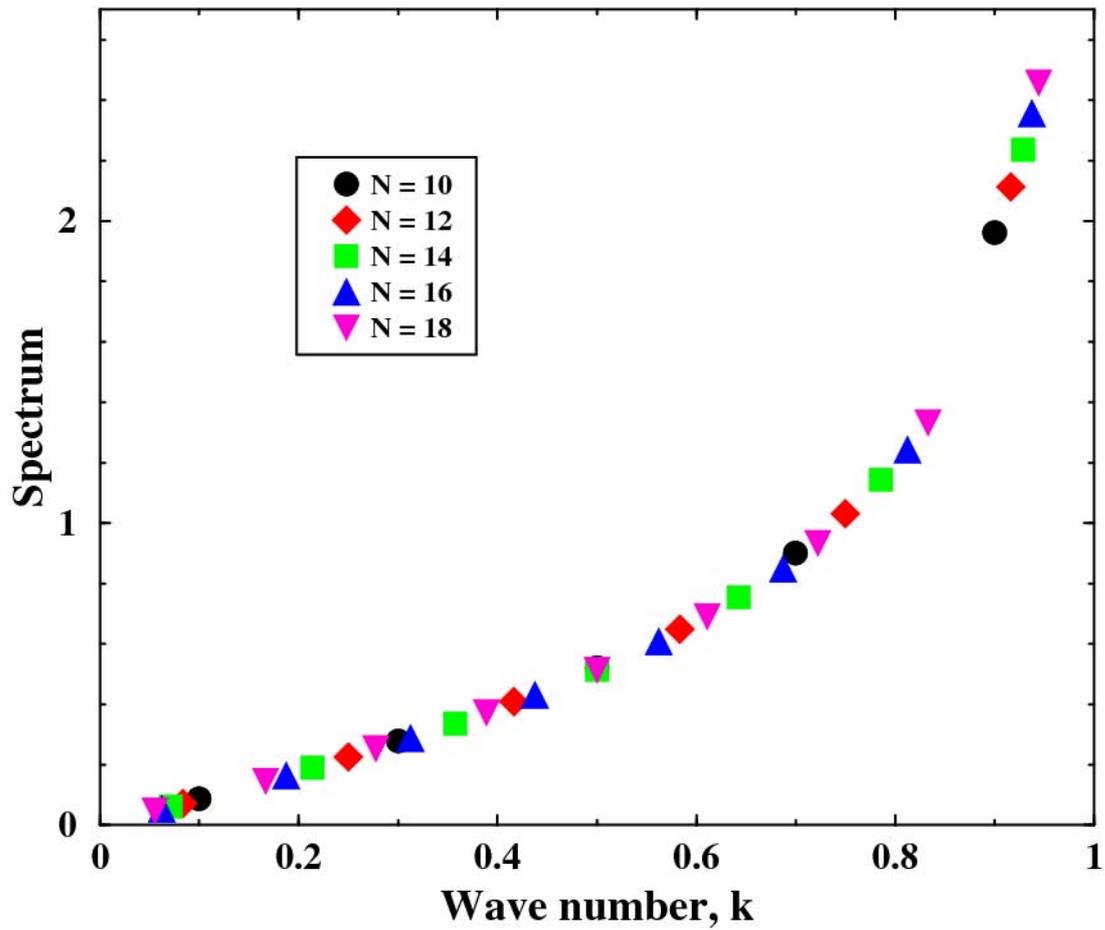

**Figure 14:** Summed spectrum of the spin waves excited by a tunnelling electron in a ring of anti-ferromagnetic atoms as a function of their wave number, *k*. The wave number is expressed in units of π/a, where a is the lattice spacing. Results for different numbers of atoms in the ring are presented: see insert.